\documentclass[Review]{TRR}

\usepackage{moreverb,url}
\usepackage{multirow}

\usepackage[colorlinks,bookmarksopen,bookmarksnumbered,citecolor=red,urlcolor=red]{hyperref}

\newcommand\BibTeX{{\rmfamily B\kern-.05em \textsc{i\kern-.025em b}\kern-.08em
T\kern-.1667em\lower.7ex\hbox{E}\kern-.125emX}}

\def\volumeyear{2021}

\begin{document}

\renewcommand{\normalsize}{\fontsize{12}{14pt}\selectfont}

\runninghead{Uribe-Laverde and Oquendo-Pati\~no}

\title{The Critical Role of the Docking Bay Assignment in the Performance of a Bus Rapid Transit System}

\author{M.A. Uribe-Laverde\affilnum{1} and W.F. Oquendo-Pati\~no\affilnum{2}}

\affiliation{\affilnum{1} Grupo de Investigaci\'on en F\'{\i}sica y Matem\'aticas Aplicadas, Faculty of Engineering,, Universidad de La Sabana, Campus Universitario Puente del Común, Km. 7 Autopista Norte, 250001 Ch\'{\i}a, Colombia\\
\affilnum{2} Grupo Simulaci\'on de Sistemas F\'{\i}sicos, Departamento de F\'{\i}sica, Facultad de Ciencias, Universidad Nacional de Colombia - Sede Bogot\'a, Carrera 30 No. 45-03, 111321 Bogotá, Colombia
}

\corrauth{M.A. Uribe-Laverde, miguel.uribe1@unisabana.edu.co}

\begin{abstract}
    Bus rapid transit (BRT) systems are a cost-effective public transportation solution moving millions of passengers every day. To optimize their operation, it is usual to implement macroscopic models that neglect the microscopic details of bus motion. In this work we show that the docking bay assignment (DBA) for the bus services at the stations, one of those disregarded microscopic details, has a large impact on the overall BRT performance. To evaluate the variations in the system's performance upon changes on the DBA, we have simulated the entire operation of a simplified BRT system using a cellular automaton microsimulation scheme. We find that the critical service frequency, above which bus queues appear, strongly depends on the DBA at the busiest stations in the system, leading to significant performance differences. Smaller critical frequencies, and poorer performance, are observed as the service demand for the services sharing a docking bay increases. By approximating the frequency optimization problem,  we show that, due to the limitations imposed to the optimal frequencies by the critical frequency, the service frequencies that minimize the total cost depend on the DBA at the busiest stations. This correlation becomes stronger as the passenger demand increases and as more importance is given to the passenger time in the total cost. Our results suggest that the frequency optimization problem in BRT systems must include the docking bay assignment at the busiest stations as an additional operation variable.
\end{abstract}

\maketitle
    \noindent A Bus Rapid Transit (BRT) system consists of dedicated corridors, typically with a length of several kilometers, where buses can move between stations without the interference of private vehicles. Additional features of BRT systems often include platform-level boarding, off-board fare collection, and busway and station alignment to the center of the street \cite{levinson_bus_2002,wright_brt_2017}. As a result, BRT systems offer a high-capacity and cost-effective public transportation alternative to the more expensive railway-based systems \cite{basso_efficiency_2019,ko_determinants_2019}. Although BRT-type systems were proposed as early as the 1930s, it is only in recent decades that these systems have been increasingly implemented in several major cities in the world like Sao Paulo, Rio de Janeiro, Bogot\'a, Istanbul, Guangzhou or Lima \cite{global_brtdata_global_2020}.

    A key feature of BRT systems is the implementation of so-called \textquotedblleft limited-stop\textquotedblright\ services that skip some stations and thus have larger average speeds~\cite{hart_methodology_2016}. As a result, most stations are served by multiple bus services. To prevent the formation of bus queues, BRT stations usually consist of multiple docking bays, where buses can stop to board and alight passengers. To provide guidance to passengers and bus drivers, at each station each bus service is assigned a specific docking bay to stop. In larger systems, however, it is usual that at the busiest stations the number of bus services is larger than the available docking bays. In those cases, there are multiple docking bay assignment (DBA) possibilities.

    The problem of finding the optimal performance parameters of a BRT system is extremely complicated due to the large number of operation variables. Published research on this field is scarce and methods to model and optimize the behavior of a BRT system have become available only in the last decade. These methods typically incorporate the definition of a global objective function that considers both competing drivers in the BRT operation, the cost in time for the users and the operation cost, at a macroscopic level. This global function, which is non-linear, is subsequently minimized subjected to fleet availability constrains. In this way, methods to find the optimal distance between stations \cite{cheng_bi-level_2019}, to optimize the bus headway \cite{galindres_guancha_asignacion_2017,koehler_real-time_2019,seman_headway_2019,suman_mitigation_2019}, and to find optimal sets of bus services and frequencies \cite{martinez_and_2017,ruano-daza_multiobjective_2018,soto_new_2017,walteros_hybrid_2015,yi_optimal_2016,zhong_optimization_2016} have been proposed.

    Owing to the global or macroscopic approach of the previous models, a limitation arises in these methods: the interaction between buses is neglected. As a consequence, the delays introduced by the formation of bus queues and the effects of the DBA at each station on the passenger flow and on the system's performance are disregarded. Mathematical models have been proposed to implement the delays due to bus queues at stations. However, these models have been used mainly to simulate and characterize the bus behavior at stations \cite{pena_optimising_2016,tirachini_bus_2011}, they have not been introduced into a global system simulation scheme. While neglecting the bus interactions might be valid for BRT systems with a small passenger load and a small bus frequency, this approach clearly does not apply for systems implemented in densely populated areas like Bogot\'a, Colombia or Rio de Janeiro, Brazil, where the peak load can be as large as 49000 and 65400 passengers per hour per direction, respectively \cite{global_brtdata_global_2020}, and where long bus and passenger queues at some stations are observed during an important fraction of the operation. Indeed, it has been established that for 121 systems in 12 countries the average bus frequency is around 116 bus/hour/direction, which is significantly high \cite{hensher_drivers_2014}. Moreover, it has been established that the bottleneck of BRT systems is at the busiest docking bays, where queues of buses appear and affect the bus flow of the entire system \cite{cain_applicability_2006,pena_optimising_2016,wright_brt_2017}. Being local in their origin, but having system-wide repercussions, queuing phenomena should be addressed by microscopic models.

    We present here a different approach to model BRT systems based on cellular automaton (CA) microsimulation. This approach allows us to study the global system's performance, while successfully accounting for queuing phenomena and the effects of changing the DBA at stations. In the microsimulation, each bus evolves according to a set of well-established rules that depend on its environment. Hence, bus interactions are inherent to the model and are properly considered. The use of CA models to simulate traffic was first proposed in 1992 by Nagel and Schreckenberg \cite{nagel_cellular_1992}. With time, more elements have been introduced into the model to simulate complex driving behavior and allow for multiple-lane highways \cite{chmura_simple_2014,maerivoet_cellular_2005,nagel_two-lane_1998,qian_cellular_2017,ruan_improved_2017,yamamoto_velocity_2017}.

    In the field of BRT systems, microsimulation methods have been implemented to model the effects of the distance between stations in a single-service corridor \cite{ancora_microsimulation_2012}, study the passenger accumulation and waiting times at bus stations \cite{rojas-galeano_discrete_2014,tomoeda_information-based_2007,larrain_danger_2020}, and study prioritization strategies at intersections  \cite{yang_performance_2013}. Commercially available microsimulation packages have also been used to evaluate the feasibility of BRT corridors and their impact on traffic~\cite{iswalt_innovative_2011,li_planning_2009}, model the bus behavior at BRT stations \cite{lin_combinatorial_2014,filipe_transmilenio_2014,kiani_mavi_bus_2018,widanapathiranage_modelling_2015} and study the emissions of a BRT corridor \cite{kim_evaluating_2019}.

    Nevertheless, to the best of our knowledge, a microsimulation-based study focused on the effects of the DBA at the stations on the performance of a BRT system has not yet been published. As we will show, these effects are important and should not be neglected. Depending on the DBA at the busiest stations, not only significant performance changes can be observed but also different optimal frequency configurations can be obtained.
\section{Simulation method}\label{Model}
    In this section we describe the cellular automaton (CA) microsimulation model for the bus movement, the user behavior model and, finally, we provide details on how the simulations were performed.

    \subsection{The cellular automata (CA) microsimulation model}
\begin{figure*}[t!]
    \centering
    \includegraphics[width=0.8\textwidth]{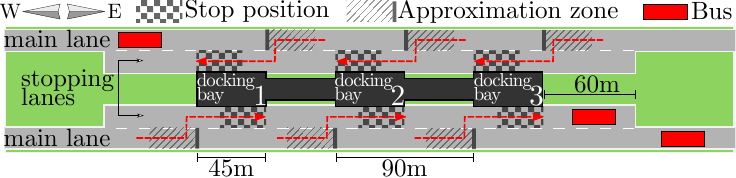}
    \caption{Schematic diagram of the stations in the simulated BRT system. Each station consists of three docking bays enumerated in the west-east direction. The buses stop at the position marked by the checker pattern. Buses can only change from the main to the stopping lane while located at the approximation zones (stripe patterned). The red arrows depict the expected approximation path of a bus.\label{system}}
    \end{figure*}

    We have created a bus corridor with a total of 46 stations stretching along the west-east direction, this is about the size of the main corridor of Transmilenio, the BRT system in Bogot\'a, Colombia. Space is discretized with cells of constant length $\delta x=3\,\mathrm{m}$. Time is also discrete, and each time-step lasts $\delta t=1\,\mathrm{s}$. The stations are uniformly distributed along the corridor with a spatial periodicity of $235\,\delta_x=705\,\mathrm{m}$. The structure of the stations is shown in Figure~\ref{system}, it is inspired by the Transmilenio stations. Each station has three docking bays in each direction, these are the platforms where buses stop. Separated by pedestrian corridors, the periodicity of the docking bays is set to $30\,\delta_x=90\,\mathrm{m}$, respectively. As also shown in Figure~\ref{system}, in the region between stations the corridor consists of a single main lane. However, to allow buses to stop without disrupting the limited-stop services, in the vicinity of the stations a stopping-lane is introduced. This lane configuration is also inspired by Transmilenio. Although the basic configuration shown in Figure~\ref{system} is enough to capture the complex response of a BRT system, it should be noted that any particular station or lane arrangements, as well as other elements such as traffic lights, could easily be introduced into the model.

\begin{figure*}[t!]
    \centering
    \includegraphics[width=0.8\textwidth]{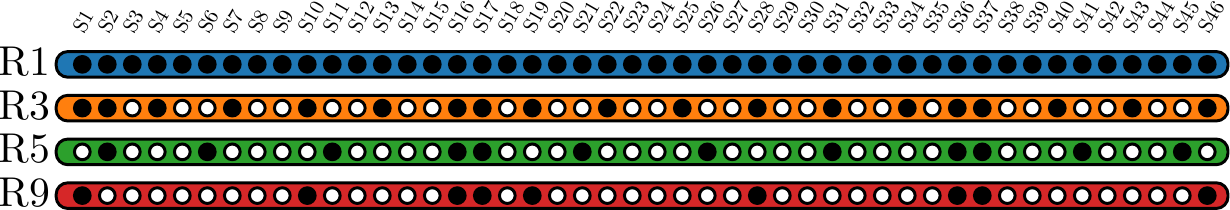}
    \caption{Bus services in operation in the BRT system. A filled station represents a stop. Service R1 is a regular service that stops at every station. Services R3, R5 and R9 are limited-stop services, which skip most of the stations in the system.\label{lines}}
    \end{figure*}

    As shown in Figure~\ref{lines}, four different bus services have been implemented, all services have the same stops in the east-bound and west-bound directions. Service R1 is a regular service that stops at every station. Services R3, R5 and R9 are limited-stop services that stop approximately every three, five, and nine stations, respectively. Additional stops have been added to make stations S16, S17, S36 and S37 the main hubs of the system, where all services make a stop. This is a typical feature of BRT systems to satisfy the demand.

    Buses are introduced in the system as a set of 10 adjacent occupied cells, therefore their length is $10\,\delta_x=30\,\mathrm{m}$  and emulates the articulated buses of Transmilenio. The rules controlling the forward motion of the buses are the usual for the Nagel and Schreckenberg (NaSch) model and can be summarized in three steps \cite{nagel_cellular_1992,schreckenberg_discrete_1995}:
    \begin{subequations}\label{accrules}
    \begin{itemize}
    \item Acceleration and braking:
    \begin{equation}\label{ar_1}
    v_i^t=\min\{v_i^{t-1}+1,g_{i}^{t}/\delta_t, v_{\rm{max}}\}.
    \end{equation}
    \item Randomization:
    \begin{equation}\label{ar_2}
    {\rm{If\ }}\xi_i^t<p{\rm{, then\ }}v_i^t=\max\{0,v_i^t-1\}.
    \end{equation}
    \item Movement:
    \begin{equation}\label{ar_3}
    x_i^t=x_i^{t-1}+v_i^t\delta_t.
    \end{equation}
    \end{itemize}
    \end{subequations}

    In rule~\eqref{ar_1}, $v_i^t$ corresponds to the velocity of bus $i$ in time-step $t$, $g_{i}^t$ corresponds to the free space in the forward direction for bus $i$ in time-step $t$, and $v_{\rm{max}}=7\,\delta_x/\delta_t=75.6\,\mathrm{km/h}$ corresponds to the maximal bus speed. According to this rule, buses are always increasing their speed unless they reach the maximal speed or they have an obstacle in front. Rule \eqref{ar_2} introduces a random braking component that is necessary to reproduce realistic phenomena when the vehicle density is large~\cite{maerivoet_cellular_2005}. At each time-step, and for every bus, a random number $0\leq \xi_i^t<1$ is generated. If this number is smaller than a random braking probability, $p=0.25$, the bus decreases its velocity by one. As a consequence of rule~\eqref{ar_2}, the cruise speed of the buses in the NaSch model is reduced to~\cite{maerivoet_cellular_2005}
    \begin{equation}\label{vns}
    v_{\rm{ns}}=v_{\rm{max}}-p\,\delta_x/\delta_t=6.75\,\delta_x/\delta_t=72.9\,\rm{km/h}.
    \end{equation}
    Besides modifying the cruise speed, and owing to the low bus density, the value of $p$ does not affect our results. Its actual value in a real BRT system depends on the driving behavior and on the road conditions. Finally, in rule \eqref{ar_3} the bus advances by a number of cells $v_i^t$ and it is now located at cell $x_i^t$. It should be clarified that $x_i^t$ points to the cell where the head of the bus is located, regardless of its direction.
    \begin{figure}[t!]
    \centering
    \includegraphics[width=0.45\textwidth]{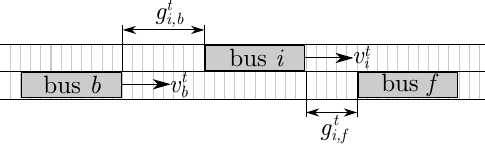}
    \caption{Definition of gaps $g_i^b$ and $g_i^f$, and the speeds $v_i$ and $v_b$. These quantities are taken into account when bus $i$ attempts to change lane according to the rule defined by Equation~\ref{cr_1}.\label{gaps}}
    \end{figure}

    The bus lateral motion is only allowed in the region where two lanes are present, the lane change is assumed instantaneous. We have set the following two criteria to be fulfilled by a bus to change lane:
    \begin{itemize}
    \item Safety:
    \begin{equation}\label{cr_1}
    v_{b}^t<g_{i,b}^t /\delta_t{\rm{\ and\ }}v_i^t <g_{i,f}^t/\delta_t,
    \end{equation}
    where bus $b$ is the bus behind bus $i$, in the opposite lane; $g_{i,b}^{t}$ is the gap between bus $i$ and bus  $b$; and $g_{i,f}$ is the gap between bus $i$ and the bus in front, in the opposite lane. A schematic definition of all gaps is shown in Figure~\ref{gaps}.
    \item Willingness:\\
    A bus in the main lane is willing to change lanes when it is located in the approximation zone of its next stop. Depicted as the stripe-patterned regions in Figure~\ref{system}, approximation zones span over $15\,\delta_x=45\,\mathrm{m}$ and are located $15\,\delta_x=45\,\mathrm{m}$ upstream of the corresponding docking bay. On the other hand, a bus in the stopping lane is willing to change back to the main lane when it has departed from its docking bay and finds an obstacle ahead. This obstacle may be a bus at a different docking bay or the end of the stopping lane.
    \end{itemize}
    Although we are defining a simplified BRT system, we would like to remark that our algorithm is highly flexible and could be adapted to reproduce other lane, station, or service configurations.
    \subsection{User behavior model}
    In real systems, the entrance profile and the origin-destination matrix, which determine passenger behavior, evolve during the day and significantly differ between the morning and evening hours. To consider these two parameters as static, we have restricted our simulations to a time window between 4:00\,a.m. and 10:00\,a.m., in the morning. In practice, similar studies should be performed for the off-peak and evening peak hours.

\begin{figure}[t!]
        \centering
        \includegraphics[width=0.45\textwidth]{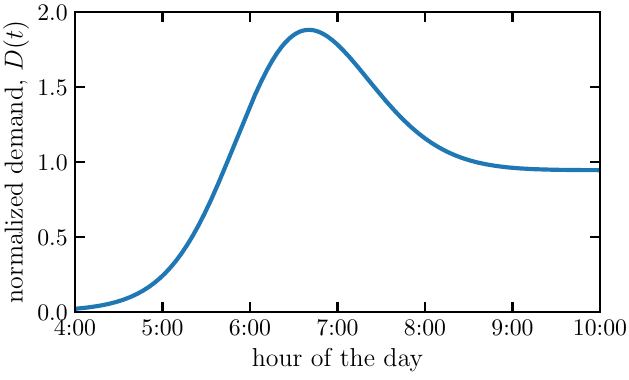}
        \caption{Normalized demand function describing the temporal-dependence of the passenger entrance rate at the BRT system between 4:00h and 10:00h. It is inspired by the real demand function of Transmilenio~\cite{transmilenio_estadisticas_2020}.\label{demandfunction}}
    \end{figure}

    Passengers are entered into the BRT system every 10 time steps. To determine the number of passengers to be inserted, we use a Poisson distribution with mean $\lambda(t)=P D(t) 10\,\delta_t/3600$. Where $P$ is the system's average passenger demand in passengers per hour, and $D(t)$ is a normalized function that describes the evolution of the passenger demand during the simulation time window. Figure~\ref{demandfunction} shows the demand function $D(t)$ used in all our simulations, which has been created using real demand profiles~\cite{transmilenio_estadisticas_2020}. The demand starts from nearly zero at 4:00\,a.m., shows a characteristic rush-hour peak close to 7:00\,a.m., and after the peak exhibits a plateau.

\begin{figure*}[t!]
        \centering
        \includegraphics[width=0.8\textwidth]{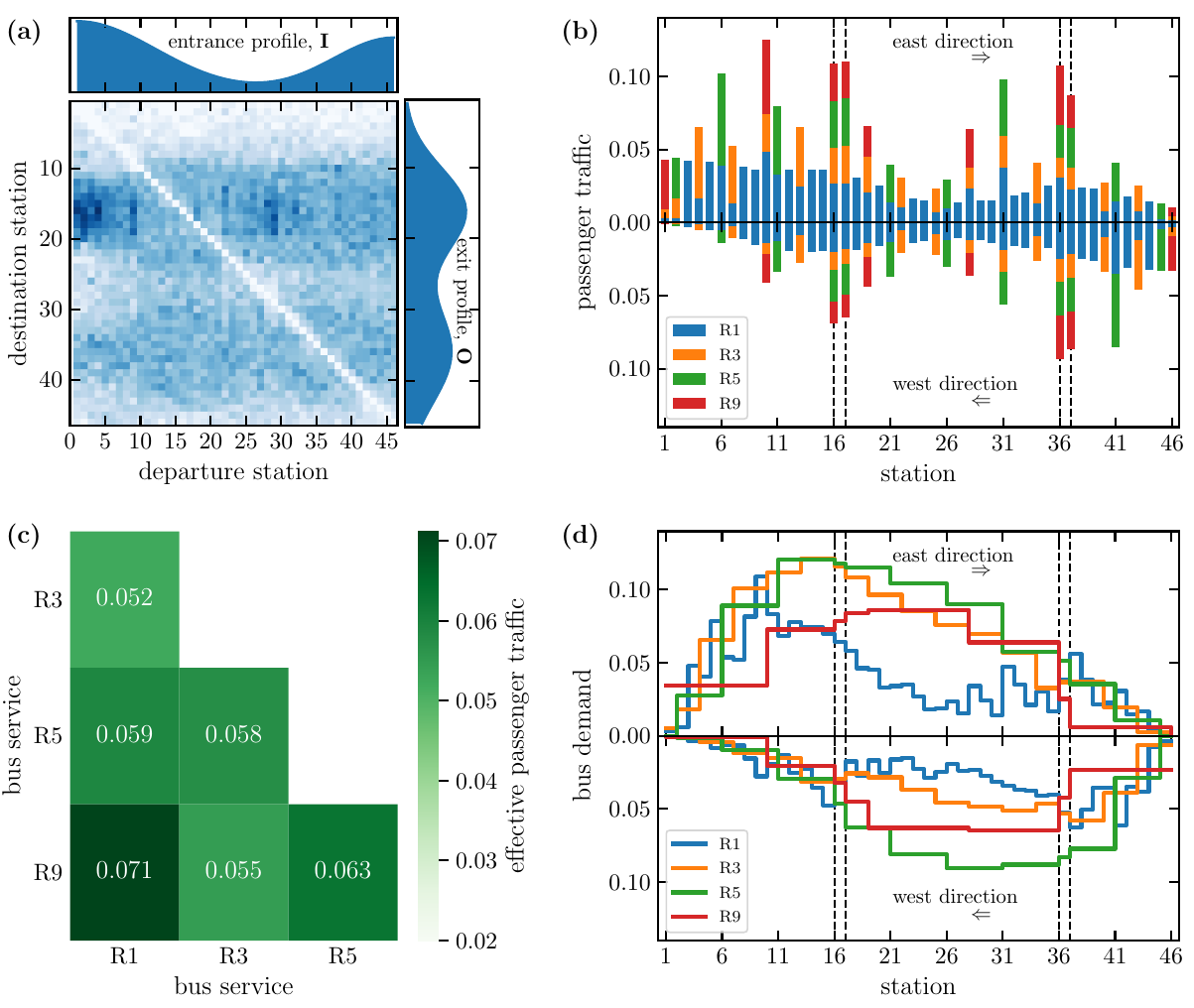}
        \caption{(a) Origin-destination matrix for the studied BRT system, $\mathbf{T}$. The entrance profile, $\mathbf{I}$, and the exit profile, $\mathbf{O}=\mathbf{TI}$, provide the probability for a passenger to enter or leave the BRT system at a given station. (b) Passenger traffic at each station. This is the probability for a passenger to board or alight a bus service at a given station in either the east or west direction. The main hubs are marked by the vertical dashed lines. (c) Effective passenger traffic for combinations of two bus services. The effective passenger traffic is calculated as the largest passenger traffic in a docking bay among the main hubs, stations 16, 17, 36 and 37. (d) Bus demand on the system. This is the probability for a passenger to be travelling in a bus service as a function of the location, either in the east or west direction. The position of the main hubs is marked by the vertical dashed lines. \label{ODmatrix}}
    \end{figure*}

    Once a passenger is created, its departure station is determined using the entrance profile vector, $\mathbf{I}$. The element $I_k$ corresponds to the probability for a passenger to enter the BRT system at station $k$. The destination station is determined by means of the origin-destination matrix, $\mathbf{T}$. The element $T_{lk}$ determines the probability for a passenger that entered the system at station $k$ to target station $l$ as its destination. The entrance profile, the origin-destination matrix, and the resulting exit profile, $\mathbf{O}=\mathbf{TI}$, that were defined for our BRT system are depicted in panel (a) of Figure~\ref{ODmatrix}. Stations 16,17, 36 and 37 have been chosen as the main hubs of the system because the exit profile peaks around these stations. 

    Given an origin-destination station pair, $(k,l)$, a passenger itinerary, $n^{k,l}$, is defined as a set of consecutive trips on different bus services such that the first trip starts at the origin station, $k$, and the last trip ends at the destination station, $j$. The weight of each passenger itinerary is defined as
    \begin{equation}
      w_{n^{k,l}} = S_{n^{k,l}} + 3 T_{n^{k,l}} + D_{n^{k,l}},
    \end{equation}
    where $S_{n^{k,l}}$ is the total number of stops, $T_{n^{k,l}}$ is the number of service transfers, and $D_{n^{k,l}}$ is the traveled distance. We have considered only passenger itineraries with up to two service changes. Since in practice passengers prefer the fastest itineraries~\cite{chriqui_common_1975}, the probability for itinerary $n^{k,l}$ to be selected by a passenger traveling from station $k$ to station $l$ decays exponentially with the weight~\cite{ulusoy_optimal_2010,ulusoy_optimal_2015}
    \begin{equation}
      P_{n^{k,l}} = \frac{e^{-w_{n^{k,l}}}}{\sum_i e^{-w_i} },
    \end{equation}
    where the sum is performed over all possible passenger itineraries joining stations $k$ an $l$ incorporating up to two service changes. After a passenger has selected its itinerary, it joins the queue for the first bus service of its travel. When a bus arrives, the passenger will board with a logistic probability that passes from one, when the bus occupation is low, to zero, when the bus is overcrowded. The midpoint of the logistic function is the nominal bus capacity, $150$ passengers, and the steepness has been set to $1\,\mathrm{pass.}^{-1}$. When the passenger finishes the trip in a bus, it either finishes its travel and leaves the BRT system or joins the queue of the next service on its way to its destination. While our itinerary selection model captures the preference of the passengers for the fastest services, it does not allow for a change of itinerary upon arrival of a suitable bus corresponding to a different itinerary. We expect that this limitation is smeared out by the multiple repetitions performed for each simulation. In addition, in real BRT systems, the itinerary changes are mostly limited to services sharing the same docking bay. Otherwise, a change in itinerary implies a fast displacement for the passenger from one docking bay to another that is seldom observed.
    
    The resulting passenger traffic, which is the probability for a passenger to board or alight a bus service at a given station, is depicted in panel (b) of Figure~\ref{ODmatrix}. In combination with the service frequency and the passenger demand, the passenger traffic is a measure of the average station occupation. As expected, the main hubs (stations 16, 17, 36 and 37) are among the busiest stations in the system. The figure also reveals an asymmetry with respect to the traveling direction, a larger number of passengers is moving in the east direction. 
    
    At each station, the passenger traffic is distributed among the available docking bays. While at most stations it is possible to assign each bus service to a different docking bay, at the main hubs at least one docking bay must bear the passenger traffic of at least two bus services. We define the \textit{effective passenger traffic} as the maximal passenger traffic at a docking bay among the main hubs. Panel (c) of Figure~\ref{ODmatrix} shows the effective passenger traffic when two bus services share a docking bay, which is the case with the most practical interest. The largest effective passenger traffic is obtained when services R1 and R9 stop at the same docking bay whereas the smallest effective passenger traffic is obtained when services R1 and R3 share the docking bay. The selection of the bus services sharing the docking bay is thus expected to have a significant influence on the occupation of the busiest docking bay at the main hubs. 
    
    The bus demand along the corridor is depicted in Panel (d) of Figure~\ref{ODmatrix}, this is the probability for a passenger to travel in each bus service as a function of the location. It can be calculated as the subtraction of the accumulated boarding and alighting probabilities. In combination with the service frequency and the passenger demand, the bus demand is a measure of the expected bus occupation. In agreement with Panel (b) of the Figure, a larger bus demand is observed in the east direction. The most demanded services in this direction are R3 and R5.

    In our simulations, the dwell time for a bus, $\tau$, is calculated every time it reaches a stop as the addition of a base dwell time of $10\,\mathrm{s}$ and a time proportional to the number of passengers alighting, $N_a$, and the passengers willing to board, $N_b$, the bus~\cite{wright_brt_2017}. In addition, an upper bound of $30\,\mathrm{s}$ has been set, thus:
    \begin{equation}\label{dwelltime}
        \tau=\min\{30\,\mathrm{s}, 10\,\mathrm{s}+0.5\,\mathrm{s}\,(N_b+N_a)\}.
    \end{equation}
    As discussed above, depending on the services sharing the docking bay different passenger traffic levels are expected at the docking bays, influencing $N_b$; on the other hand, each bus service exhibits a different bus demand, influencing $N_a$. It is therefore expected that changes in the docking bay assignment at the main hubs introduce changes in the dwell time at the busiest docking bays. Since the dwell time has a direct impact on the formation of queues, the interplay between the docking bay assignment, the service frequencies and the passenger demand may affect the performance of the system. This effect is largely neglected by macroscopic models.
    \subsection{The simulation process}\label{model_simulation}
    \begin{figure}[t!]
        \centering
        \includegraphics[width=0.4\textwidth]{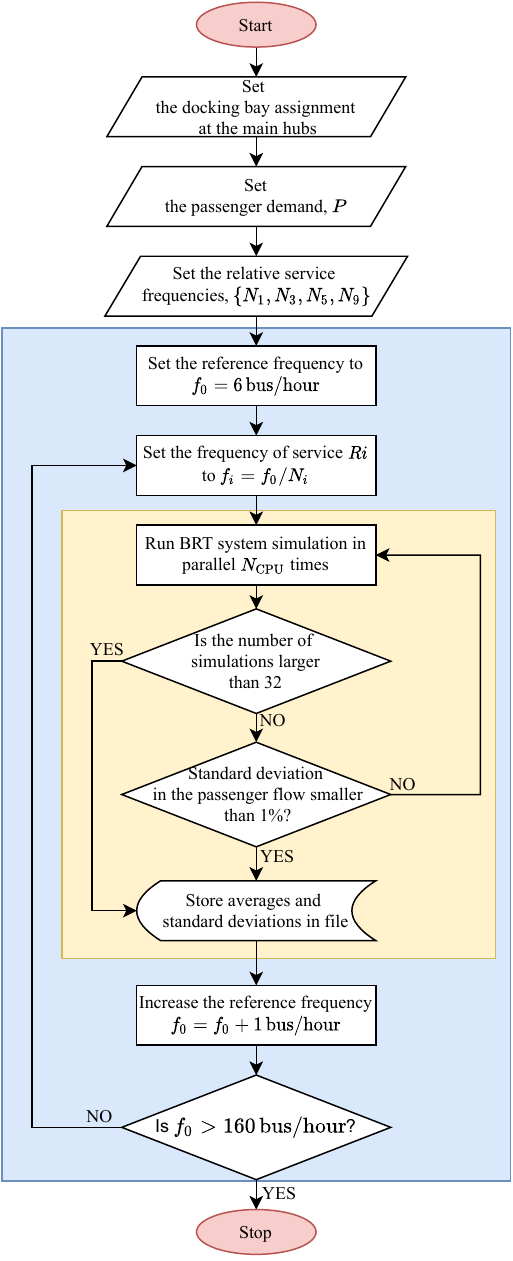}
        \caption{Flow chart for the frequency scan process. The inputs for the simulation are the docking bay assignment, the passenger demand, and the relative service frequencies. The output is a file containing the averages and standard deviations, computed over all performed simulations, of the BRT system's performance parameters for each value of the reference frequency $f_0$.\label{FlowChart}}
    \end{figure}
    The main process carried out through this research is the frequency scan process described in the flowchart presented in Figure~\ref{FlowChart}. The goal of this process is to determine the evolution with the bus service frequency of the system's performance parameters: average bus speed, average passenger speed, global passenger flow and operation cost. As shown in Figure~\ref{FlowChart}, the frequency scan process has three input parameters:
    \begin{itemize}
        \item Docking bay assignment (DBA) at the main hubs (stations S16, S17, S36 and S37). For simplicity, the same assignment is assumed at the four stations. The assignment is defined with respect to the east-bound direction and, to keep the symmetry, is inverted for buses moving to the west. If for a given service the buses moving to the east are set to stop at docking bay 1, the buses moving to the west will stop at docking bay 3. The notation of the DBA is defined as follows: if services R1, R3, R5 and R9 are assigned the docking bays 1,2,3 and 1, respectively, the DBA will be noted as [R1,R9]-[R3]-[R5]. The square brackets in this notation represent each docking bay. To reduce bus interactions, at the stations served by three or less bus services each bus service is assigned an exclusive docking bay to stop.
        \item Average passenger demand, $P$, in passengers per hour. This is equal to the total number of passengers entering the system during the simulation divided by the 6-hour operation window.
        \item Relative service frequencies. To keep a low dimensionality, during a frequency scan process the relative frequency of the bus services is kept constant. The relative service frequencies are represented by a set of four numbers \{$N_\text{1}$, $N_\text{3}$, $N_\text{5}$, $N_\text{9}$\}, where $N_i\in\{1,2,3\}$. The frequency of service R$i$ is set as:
        \begin{equation}\label{servfreq}
             f_i=f_0/N_i,\qquad \text{for } i\in \text{\{1,3,5,9\}},
        \end{equation}
        where $f_0$ is the scanned reference frequency. The larger $N_i$ is, the smaller the relative frequency of the corresponding bus service.
    \end{itemize}

    Once the input parameters are set, the reference frequency $f_0$ is scanned in the range from $6$ to $160\,\mathrm{bus/hour}$ with a step of $1\,\mathrm{bus/hour}$. For each $f_0$ value, the actual service frequencies are computed using Equation~\ref{servfreq}. Subsequently, the entire operation of the BRT system is simulated in the time window between 4:00\,a.m. and 10:00\,a.m. and the system's performance parameters are calculated. Since the model has many stochastic components, the simulation for each value of $f_0$ is launched in parallel $N_{\rm{CPU}}$ times using a different seed, with $N_{\rm{CPU}}$ the number of available logical processors (either $8$, $16$ or $24$). This process is repeated until the relative standard deviation of the passenger flow falls below $1\%$ or the number of simulations is larger than $32$. Finally, the average values and standard deviations of the bus speed, the passenger speed, the passenger flow, and the operation cost are calculated using the results of all the performed simulations. The entire simulation software has been written by the authors using the Python and C++ programming languages.

    \section{Model Validation}\begin{figure*}
    \centering
    \includegraphics[width=0.8\textwidth]{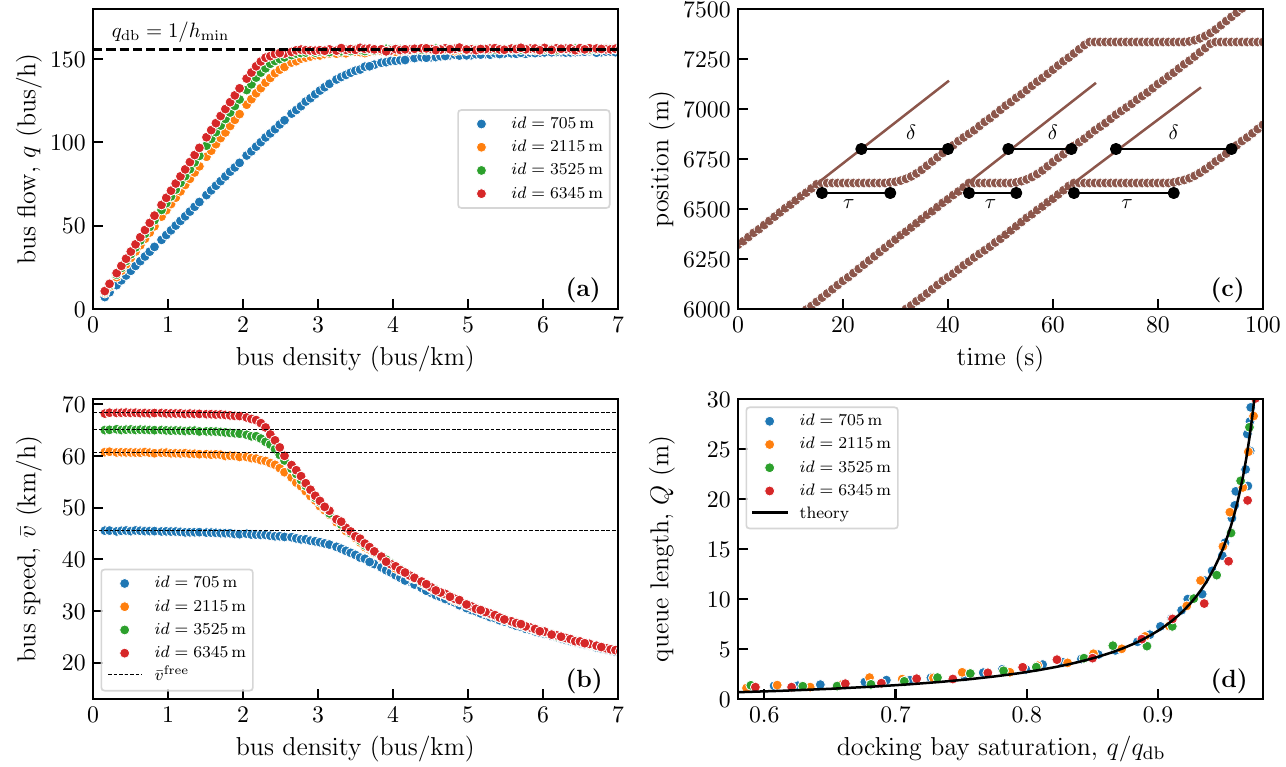}
    \caption{Model validation data obtained using an average dwell time of $\bar{\tau}=15\,\rm{s}$ in a BRT system with periodic boundary conditions. (a) Bus flow as a function of the bus density for different distances between stops, the dashed line marks the maximal bus departure rate $q_{\rm{db}}$. (b) Average bus speed as a function of the bus density for different distances between stops. The dashed lines in each case represent the expected average speed when no queues are present at the stations, calculated by means of Equation~\eqref{vfree} with $\delta=20.89\,\rm{s}$. (c) Some examples of the space-time diagram of buses reaching a station, the dwell time $\tau$ and the effective delay $\delta$ are highlighted in each case. The solid lines are an extrapolation of the bus trajectory before reaching the stop. (d) Average queue length, as obtained from the speed data using Equation~\eqref{queues}, as a function of the docking bay saturation $q/q_{\rm{db}}$. The solid line represents the expected theoretical behavior.\label{validation}}
    \end{figure*}
    To test the capacity of our CA model to reproduce some theoretically expected bus movement and queuing phenomena, we investigated the fundamental diagram (bus flow vs. bus density) in a simplified BRT system where all buses are set to stop every 1, 3, 5 or 9 stations. Before starting each simulation, buses are randomly and homogeneously distributed among the entire corridor. In addition, periodic boundary conditions are set to keep the number of buses constant. The bus flow and bus average speed are measured after the system is in a steady state. To avoid the complications of a non-uniform passenger demand, for this analysis the dwell time is randomly chosen at each stop following a Poisson distribution with mean $\bar{\tau} =15\,\rm{s}$.

    The computed fundamental diagrams for the different stop distances are shown in Figure~\ref{validation}(a). The bus flow, $q$, evolves as expected for dedicated lanes~\cite{haitao_analytical_2018}. At low bus densities it increases linearly. As the density increases, however, the bus flow saturates at the largest possible departure rate from a docking bay, $q_{\rm{db}}$, which is independent of the distance between stops. By averaging the bus flow at high bus densities we obtain  $q_{\rm{db}}=155.7(5)\,\rm{bus/h}$, which corresponds to an average minimal bus headway of $h_{\rm{min}}=1/q_{\rm{db}}=23.11(7)\,\rm{s}$.

    Figure~\ref{validation}(b) exhibits the average bus speed as a function of the bus density for the different stop distances. The reduction of the average speed with the bus density is evidence of the emergence of bus queues at the stations. Theoretically, the average bus speed can be modeled as the distance between stops divided by the total time spent between two adjacent stops:
    \begin{equation}\label{queues}
        \bar{v}_i = \frac{id}{(id-Q)/v_{\rm{ns}}+Q/(L_{\rm{b}} q_{\rm{db}})+\bar{\delta}}, \qquad i\in\{1,3,5,9\},
    \end{equation}
    where $d$ is the distance between stations, $Q$ is the average queue length at the stations, $L_{\rm{b}}$ is the bus length, and $\bar{\delta}$ is the average effective delay introduced by the bus stop. The first term in the denominator of Equation~\eqref{queues} represents the time average time a bus moves at cruise speed, $v_{\rm{ns}}$. The second term represents the average time spent queueing, the term $L_{\rm{b}} q_{\rm{db}}$ corresponds to the average bus speed while in a queue. At low bus densities, when queues are not present at the stations and buses can freely move between stops, the average speed can be calculated as
    \begin{equation}\label{vfree}
        \bar{v}_i^{\rm{free}} = \frac{id}{id/v_{\rm{ns}}+\bar{\delta}}, \qquad i\in\{1,3,5,9\}.
    \end{equation}
    From the speed data at low densities, the average effective delay can be estimated as $\bar{\delta} = 20.89(2)\,\rm{s}$ and, as expected, it does not depend on the distance between stops. The computed average bus speeds in free-flow conditions are depicted as dashed lines in Figure~\ref{validation}(b). To understand the reason why $\bar{\delta}$ is several seconds larger than the average dwell time, $\bar{\tau}$, we show some examples of the space-time diagram for buses reaching a station in Figure~\ref{validation}(c). The dwell time, $\tau$, corresponds to the time the bus is stopped. Nevertheless, the effective delay introduced by the stop at the station, $\delta$, is in all cases larger than $\tau$ due to the finite bus acceleration. We also notice that $h_{\rm{min}}$ is slightly larger than $\bar{\delta}$, this is  due to additional delays introduced by the bus interactions when queues appear. 

    According to the BRT Planning Guide, Ref.~\cite{wright_brt_2017}, the average length of the bus queues, at a docking bay when a random dwell time is introduced in the system is given by:
    \begin{equation}\label{queuelength}
        Q=\frac{L_{\rm{b}}}{2}\left(I_\text{arr}+I_\text{dep}\right) \frac{x^2}{(1-x)},
    \end{equation}
    where $I_\text{arr/dep}$ is the irregularity of arrivals/departures at the docking bay, defined as the ratio between the variance and the squared mean of the arrival/departure times, and $x=q/q_{\rm{db}}$ is the docking bay saturation level, defined as the probability of finding a docking bay occupied by a bus. In this simplified version of the BRT system, all the stations are expected to behave in the same way. Therefore, the irregularity of arrivals and departures are the same. The irregularity of departures is directly related to the random nature of the dwell time and can be calculated as:
    \begin{equation}
    I_\text{dep}=\bar{\tau} q_{\rm{db}}^2,
    \end{equation}
    where we have considered that the variance in the departures is given by the variance in the nominal dwell time: $\sigma^2_\tau=\bar{\tau}$, and that the average departure rate in the presence of queues corresponds to the minimal headway $h_{\rm{min}}=1/q_{\rm{db}}$. The average queue length in equation \eqref{queuelength} can thus be written, in terms of the bus flow, as:
    \begin{equation}\label{queue}
    Q=L_{\rm{b}}\tau\frac{q^2}{(1-q/q_{\rm{db}})}.
    \end{equation}
    In the last equation the distance between stops does not appear explicitly, the average queue length at a docking bay only depends on the bus flow and on the average departure rate. Figure~\ref{validation}(d) shows the queue length, calculated using the average speed data and the Equation~\eqref{queues}, as a function of the bus flow for all stop distances. Regardless of the distance between the stops all data collapses in a single curve that follows remarkably well the theoretical prediction given by Equation~\eqref{queue}. This observation confirms that the average bus speed can be effectively modeled by Equation~\eqref{queues} and that our model creates bus queues in agreement to Equation~\eqref{queuelength}.

    We conclude that the CA model successfully reproduces the expected bus behavior both in conditions of free flow and when queues appear at the stations.

    \section{Results and Discussion}
    In the following, we simulate the complete BRT system. Buses are incorporated into the system according to each service frequency and passengers are introduced as described in the \textit{User behavior model}, the dwell time at each stop is calculated by means of Equation~\eqref{dwelltime}.

    This section is organized as follows. First, we study the behavior of the BRT system when all the service frequencies are equal and show that, due to differences on the demand among bus services, some DBA at the main hubs are more prone to the formation of queues. Second, we introduce a cost function and show that these differences in queue formation lead to different optimal configurations. Finally, we approximate the frequency optimization problem by scanning over different relative frequency configurations and show that different DBA lead to different sets of service frequencies and to different minimal values in the cost function.
        \begin{figure*}[t!]
            \centering
            \includegraphics[width=0.8\textwidth]{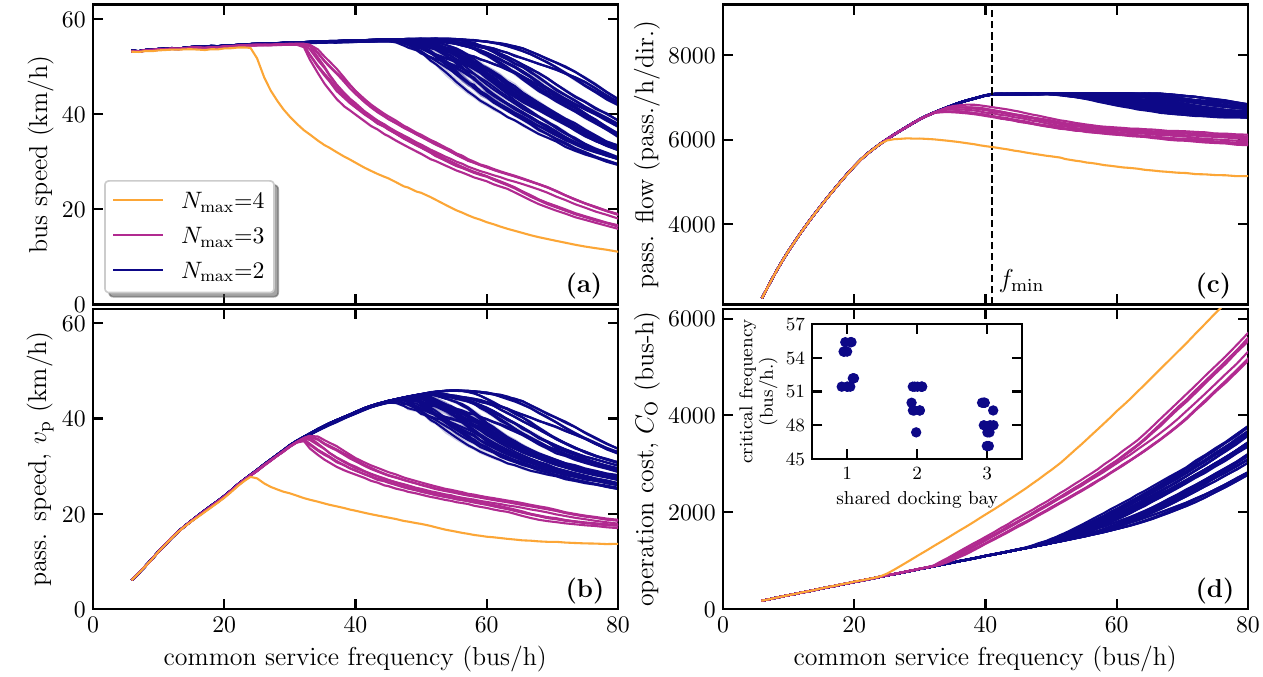}
            \caption{Dependence of the BRT system's performance parameters on the common bus service frequency for 51 different docking bay assignments at the main hubs with an average passenger demand $P=4.0\times 10^4\,\mathrm{pass./hour}$. The colors represent the maximum number of services sharing the same docking bay, $N_{\mathrm{max}}$, and it is the same for all panels. (a) Average bus speed. (b) Average passenger speed. (c) System-wide average passenger flow. (d) Operation cost. The standard deviation is represented by a shadowed region around the data and is barely visible. The vertical line in panel (c) marks the position of $f_{\rm{min}}$, the minimal frequency required to satisfy the demand. The inset in panel (d) shows how the critical frequency  changes depending on the docking bay being shared for all the DBA with $N_{\mathrm{max}}=2$. \label{stopDependence_2}}
        \end{figure*}
        \begin{figure*}[t!]
            \centering
            \includegraphics[width=0.8\textwidth]{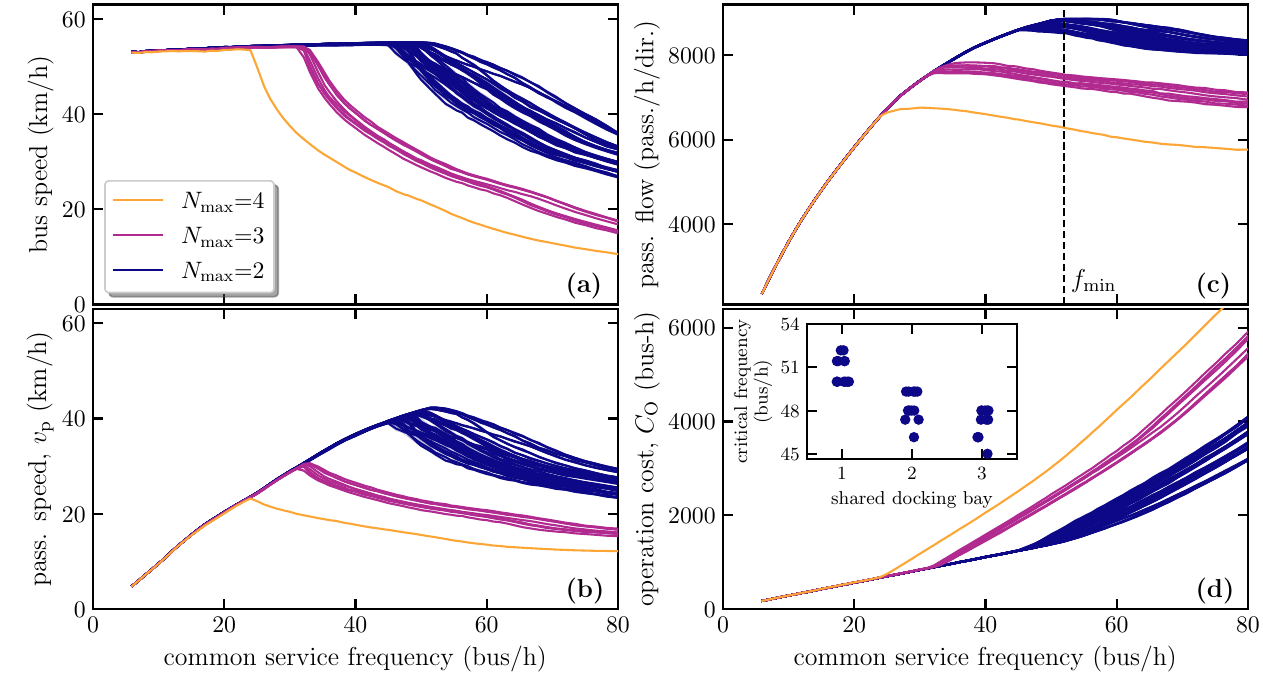}
            \caption{Dependence of the BRT system's performance parameters on the common bus service frequency for 51 different docking bay assignments at the main hubs with an average passenger demand $P=5.0\times 10^4\,\mathrm{pass./hour}$. The colors represent the maximum number of services sharing the same docking bay, $N_{\mathrm{max}}$, and it is the same for all panels. (a) Average bus speed. (b) Average passenger speed. (c) System-wide average passenger flow. (d) Operation cost. The standard deviation is represented by a shadowed region around the data and is barely visible. The vertical line in panel (c) marks the position of $f_{\rm{min}}$, the minimal frequency required to satisfy the demand. The inset in panel (d) shows how the critical frequency  changes depending on the docking bay being shared for all the DBA with $N_{\mathrm{max}}=2$.\label{stopDependence_3}}
            \end{figure*}
      \subsection{Docking Bay Assignment and Critical Frequency}
        We start our analysis by studying the effects on the Bus Rapid Transit (BRT) system's performance of variations in the docking bay assignment (DBA) at the main hubs: stations 16, 17, 36 and 37. To do this,  we performed frequency scans for a total of $51$ different DBA. This number is smaller than the $3^4=81$ possible arrangements because of the elimination of operational duplicates: for example, the DBA where all services stop at docking bay 1 is equivalent to the DBA where all service stop either at docking bay 2 or at docking bay 3. For DBA with one free docking bay, it has always been the number 3. For the DBA with two free docking bays, the used docking bay was the number 1. These restrictions reduce the number of possibilities to 51. As mentioned above, at the remaining stations an exclusive docking bay is assigned to each bus service to minimize bus interactions. For simplicity, at this point we have set the frequency of all bus services to be the same, i.e., the relative frequencies have been set to $\{N_\text{1},N_\text{3},N_\text{5},N_\text{9}\}=\{1,1,1,1\}$. This analysis has been carried out for two different passenger demands, namely $P=4.0 \times 10^4\,\mathrm{pass./hour}$ and $P=5.0\times 10^4\,\mathrm{pass./hour}$.

        The results of the frequency scans are shown in Figures~\ref{stopDependence_2} and~\ref{stopDependence_3} for passenger demands $P=4.0 \times 10^4\,\mathrm{pass./hour}$ and $P=5.0\times 10^4\,\mathrm{pass./hour}$, respectively. In both figures, the evolution of the BRT system's performance parameters with the common service frequency is presented as follows: average bus speed in Panel (a), average passenger speed in Panel (b), global passenger flow in Panel (c) and operation cost in Panel (d). For comparison purposes, the scales are the same in both figures; in addition, the color of each dataset has been assigned according to the maximal number of services sharing the same docking bay, $N_{\rm{max}}$. The standard deviation of the results shown in Figures~\ref{stopDependence_2} and~\ref{stopDependence_3} is depicted as a shadowed region around the data and is barely visible, which testifies to the reproducibility of the simulations.
          
        When the common service frequency is small, all system's performance parameters in Figures~\ref{stopDependence_2} and~\ref{stopDependence_3}  show the same evolution, regardless of the DBA. In this regime of low frequencies, bus interactions are not important. However, for each DBA there is a certain critical frequency above which all the system's performance parameters diverge from the general trend. Clearly, the value of the critical frequency for each DBA is mainly determined by $N_{\rm{max}}$. The fewer bus services that share a docking bay, the larger the value of the critical frequency. This result highlights the importance of having multiple docking bays available at the busiest stations in BRT systems. The DBA with $N_{\rm{max}}=2$, which make use of all available docking bays, yield the largest critical frequencies and are therefore the DBA of relevance in practical operation.
        
        Notably, even among DBA with the same value of $N_{\rm{max}}$ there are still significant differences in the critical frequency. The performance of the BRT system is thus strongly dependent on how the stops are arranged at the main hubs, four stations in the system. This dependence is completely overlooked by macroscopic models where bus interactions and the influence of the passenger demand on the dwell time are ignored.

        The origin of the critical frequency can be understood by looking at the average bus speed curves shown in Panel (a) of Figures~\ref{stopDependence_2} and~\ref{stopDependence_3}. For frequencies below the critical frequency, the bus speed is seen to slightly increase. This is a consequence of reduced passenger load at stations, hence reduced dwell times. Above the critical frequency, however, the bus speed starts to rapidly decrease as the common service frequency increases. This is evidence of the formation of bus queues. Therefore, for each  set of  relative frequencies and for each DBA, the critical frequency can be defined as the value of the reference frequency above which queues start to form in the system. By comparing data in Figures~\ref{stopDependence_2} and~\ref{stopDependence_3} we can conclude that the critical frequency is reduced as the passenger demand increases. This is a consequence of longer dwell times and shows that the actual value of the critical frequency is the result of the interplay between bus interactions and service demand at the stations.

        Panel (b) in Figures~\ref{stopDependence_2} and~\ref{stopDependence_3} show the evolution with the common service frequency of the average passenger speed. The speed for an individual passenger is calculated as the distance between the origin and destination station divided by the total time spent in the system. For passengers still in the system at the end of the simulation the speed is calculated as the distance between the origin station and their final location divided by the time spent in the system. Below the critical frequency, the passenger speed increases with the service frequency. However, above the critical frequency of each DBA the passenger speed starts to decrease owing to the ever-increasing times spent in the bus queues. As evidenced by comparing the data in Figures~\ref{stopDependence_2} and~\ref{stopDependence_3}, the average passenger speed is reduced as the passenger demand increases because of longer waiting times at the stations and longer dwell times. 
        
        Panel (c) in Figures~\ref{stopDependence_2} and~\ref{stopDependence_3} shows the evolution of the passenger flow with the common service frequency. This is a key parameter as it measures the capacity of the system to transport passengers. At low frequencies, the passenger flow rapidly increases with the common bus service. In general, the passenger flow saturates and reaches a plateau at a certain frequency, $f_{\rm{min}}$, marked as a vertical line in the figure; for the set of chosen relative frequencies, this is the minimal reference frequency required to properly satisfy the passenger demand. As expected, $f_{\rm{min}}$ increases with the passenger demand. Within the passenger flow plateau, the passenger demand is properly satisfied: no accumulation of passengers occurs at the stations while bus queues do not form. In practice, it is expected for BRT systems to operate within this region.

        Remarkably, it becomes clear from the data presented in Panel (c) of Figures~\ref{stopDependence_2} and~\ref{stopDependence_3} that if the critical frequency of a given DBA is smaller than $f_{\rm{min}}$, the DBA never reaches the passenger flow plateau and falls short of satisfying the passenger demand. This is the case for all DBA with $N_{\rm{max}}=4$ and $N_{\rm{max}}=3$. As the passenger demand increases, $f_{\rm{min}}$ increases whereas the critical frequency of all DBA decreases; therefore, the number of DBA able to properly satisfy the demand is reduced when the passenger demand increases from $P=4.0 \times 10^4\,\mathrm{pass./hour}$ to $P=5.0\times 10^4\,\mathrm{pass./hour}$. Only by implementing methods that properly consider the bus interactions and the service demand can these optimal DBA be determined.
        
        For each DBA reaching the passenger flow plateau, the upper bound of the plateau is the critical frequency. For frequencies above this value the passenger flow starts to decrease, evidencing that the formation of bus queues in a BRT system negatively affects its capacity to transport passengers.

        Panel (d) in Figures~\ref{stopDependence_2} and~\ref{stopDependence_3} shows the evolution of the operation cost, $C_{\rm{O}}$, expressed in bus-hour. This unit represents the cost of operation of one single bus during a one-hour period. At low frequencies, the operation cost increases linearly with the service frequency, regardless of the DBA. Nevertheless, above the critical frequency a steeper dependence with the service frequency is observed. This is because more buses are needed to sustain the operation when queues are present in the BRT system. 
        
        The insets in panel (d) depict the critical frequency, measured as the frequency where the passenger speed reaches a maximum, as a function of the docking bay being shared for the DBA with $N_{\rm{max}}=2$. Notably, regardless of the passenger demand, larger critical frequencies are observed when the shared docking bay is the number 1 (number 3 for west-bound services). Queues are thus less likely to form when the busiest docking bay is the first one that the buses find when reaching a station. The origin of this difference lies in the fact that, while buses stopping at docking bay 1 do not have to worry about any traffic in the stopping lane when approaching their stop, buses stopping at docking bays 2 or 3 may interact with buses already in the stopping lane and may thus experience a more congested arrival.

        Nevertheless, the docking bay being shared is not the only determinant of the critical frequency. As seen in the insets of panel (d) in the Figures~\ref{stopDependence_2} and~\ref{stopDependence_3}, even when the same docking is being shared different critical frequencies are observed. Since all services have the same frequency, the origin of these differences can only be explained in terms of differences in the average dwell time. After a closer look at the data, we note that the largest critical frequencies are obtained when services R1 and R3 share the same docking bay. As depicted in panel (c) of Figure~\ref{ODmatrix}, this combination exhibits the smallest effective traffic at the main hubs. With fewer people boarding and alighting at the main hubs, shorter dwell times are expected for these DBA. 
        
        We note, however, that the smallest critical frequencies are consistently obtained when services R3 and R5 are the ones sharing a docking bay, which does not correspond to the combination with the largest combined traffic according to panel (c) of Figure~\ref{ODmatrix}. The reason of this apparent discrepancy lies in the bus occupation. As shown in panel (d) of Figure~\ref{ODmatrix}, services R3 and R5 are the ones with the largest bus demand at the main hubs. If buses are crowded the boarding probability decreases and more passengers are expected to be waiting at the stations, the expected dwell time is thus increased. The differences in the critical frequencies among the studied DBA are thus driven by both the number of passengers waiting for the buses and the bus occupation, at the main hubs.
        
        Despite the imposed restrictions, which significantly reduced the dimensionality of the system, the results shown in Figures~\ref{stopDependence_2} and~\ref{stopDependence_3} offer valuable information regarding the operation of a BRT system. We have shown that, due to bus interactions and differences in the service demand, the performance of the BRT system is significantly affected by changes in the DBA at the main hubs. Some DBA are more prone to the formation of bus queues than others. We also showed that the presence of bus queues negatively affects the performance of the system not only from the passengers' point of view but also from the perspective of the operation cost. A BRT system where queues form with regularity is therefore susceptible to significant performance improvements. Our data also suggest that as the passenger demand increases fewer docking bay assignments exhibit a passenger flow plateau and are thus able to satisfy the demand. Only with methods where the interaction between buses and the bus demand are properly considered can the limits of the passenger flow plateau for a given DBA be accurately determined.
        
        \subsection{Total Cost Function and Optimal Frequencies}\begin{figure*}[t!]
            \centering
            \includegraphics[width=0.9\textwidth]{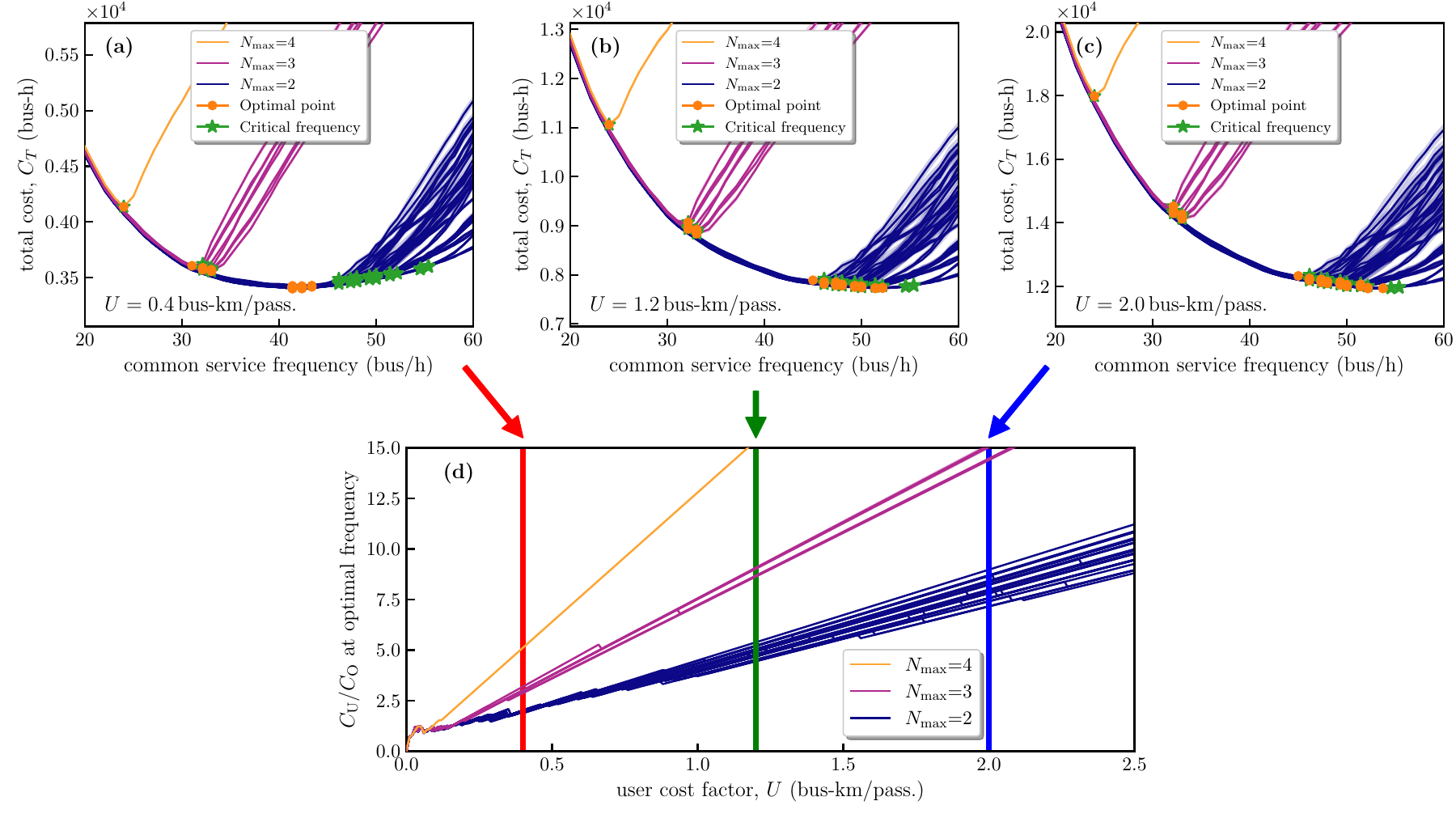}
            \caption{Evolution of the total cost of the BRT system with the common service frequency for different DBA at the main hubs. The user cost factor has been set to: (a) $U=0.4\,\mathrm{bus-km/pass.}$, (b) $U=1.2\,\mathrm{bus-km/pass.}$ and (c) $U=2.0\,\mathrm{bus\-km/pass.}$. Each line represents a DBA and the line color is assigned according to the maximum number of services sharing a docking bay. For each DBA, the orange points and green stars represent the optimal frequency, for which the total cost is minimized, and the critical frequency. Panel (d) shows the user cost to operation cost ratio at the optimal frequency as a function of the user cost factor. In all cases the passenger demand has been set to $P=4.0\times 10^4\,\mathrm{pass./h}$. The standard deviations are represented by a shadowed region around the data and are barely visible.\label{stopDependence-totalcost}}
            \end{figure*}

        To find optimal operating conditions in BRT systems, a total cost function is usually defined and minimized. This function considers the two competing drivers in BRT operation: the operation cost and the cost in time for the users. We define the total cost, $C_{\rm{T}}$, as
        \begin{equation}\label{perfEq}
            C_{\rm{T}}=C_{\rm{O}}+C_{\rm{U}}=C_{\rm{O}}+U\frac{6\,\mathrm{h}\, P}{v_{\mathrm{p}}}.
        \end{equation}
        The operation cost, $C_{\rm{O}}$, is directly obtained from our simulations and is measured in bus-h. In a real system, the bus-h should include the cost of personal, fuel, maintenance, and other related costs per bus per hour of operation. The user cost $C_{\rm{U}}$ is proportional to the total number of passengers, 6\,h\,$P$, and is inversely proportional to the average passenger speed, $v_{\mathrm{p}}$. It quantifies the time a passenger spends in the system per traveled distance unit, including the waiting times at the stations. The weight of this term in real BRT systems is related to the importance that the system operators give to the user's time. Since in this work we are not interested in absolute costs, only one weight parameter is introduced in the total cost function in Eq.~\ref{perfEq}, this is the user cost factor, $U$. It measures the cost, in bus-hour, given to the average time in hours that a passenger spends to move one kilometer in the system, it is expressed in bus-h\,km/h/pass, a rather long unit which in the following will be referred to as bus-km/pass.

        Panels (a), (b) and (c) in Figure~\ref{stopDependence-totalcost} show the evolution of the total cost with the common service frequency for all the studied DBA, and for three different values of the user cost factor: $0.4$, $1.2$ and $2.0\,\rm{bus\mbox{-}km/pass}$. The passenger demand in all cases is $P=4.0\times 10^4\,\mathrm{pass./h}$. As seen in all panels, for each DBA and user cost factor values, the total cost exhibits a minimum at a certain optimal frequency marked with an orange point in the figure. The lowest total costs are obtained for the DBA with $N_{\rm{max}} = 2$, evidencing the practical advantage of using all the available docking bays.
        
        In principle, the optimal service frequency should be independent of the DBA at the main hubs and should only depend on the service demand and the user cost factor. However, as seen in Panels (a), (b) and (c) in Figure~\ref{stopDependence-totalcost}, among the studied DBA different optimal frequencies, and different minimal total costs, are observed. Furthermore, these differences become larger as the user cost factor increases. This is a consequence of the limitation imposed to the optimal frequency by the formation of bus queues, i.e., by the critical frequency (green stars in Figure~\ref{stopDependence-totalcost}). As more importance is given to the user cost, the optimal frequency tends to move towards higher values to reduce the passengers' waiting times at the stations. However, it cannot be larger than the critical frequency, above which bus queues appear and the total cost exhibits a sharp increase. Therefore, it is expected that as the user factor increases fewer DBA can successfully minimize the total cost without the restrictions imposed by the formation of bus queues. For the remaining DBA the optimal frequency must be equal to the critical frequency. This effect is clearly portrayed in Panels (b) and  (c) in Figure~\ref{stopDependence-totalcost}. 
        
        Although queues are avoided in the optimal configurations, the observed differences in the optimal frequency and total cost among the studied DBA are a direct consequence of bus interactions and can only be studied by microscopic models.

        A natural question appears at this point regarding the values that the user cost factor, $U$, can take in real systems. According to previous case studies carried out in New Jersey (United States)~\cite{ulusoy_optimal_2015}, Beijing (China)~\cite{chen_optimization_2015}, and Chengdu (China)~\cite{qu_optimizing_2016}, the user cost to operation cost ratio in optimal operating conditions can be as small as 2 and as large as 11. These ranges are also confirmed by theoretical studies based on real data for Santiago (Chile) and London (England)~\cite{basso_efficiency_2014,basso_efficiency_2019}. 
        The calculated user cost to operation cost ratio, $C_{\rm{U}}/C_{\rm{O}}$, in optimal operating conditions for all the studied DBA in the case with $P=4.0\times 10^4\,\mathrm{pass./h}$ is shown in  Panel (d) of Figure~\ref{stopDependence-totalcost}. It becomes clear that within the range between $0$ and $2.5\,\mathrm{bus\mbox{-}km/pass.}$ the user cost to operation cost ratio is within the aforementioned range for all the DBA with $N_{\rm{max}}=2$, which are the configurations of practical relevance. Therefore, the observed performance differences between different DBA are feasible and not the consequence of unrealistically high user cost factors.

    \subsection{Best Service Frequencies} 
        \begin{table*}
        \small\sf\centering
        \caption{Best service frequency configurations and the corresponding total cost, user cost to operation cost ratio, and average passenger speed, obtained for different values of the passenger demand, $P$, the user cost factor, $U$, and the docking bay assignment (DBA) at the main hubs. The DBA are noted as A: [R1,R3]-[R5]-[R9], and B: [R3,R5]-[R1]-[R9] and C: [R1]-[R9]-[R3,R5]. For each configuration, the best frequencies for services sharing a docking bay are highlighted in bold. The last column,  $\Delta C_{\rm{T}}$, shows the increment in the total cost if, while using the best service frequencies for DBA A, the DBA at the main hubs is changed to configurations B or C. The parentheses after a quantity represent the standard deviation in the last significant figure.}
        \begin{tabular}{|c|c|p{1.0cm}||c|c|c|c|c|c|c|c|c|}
            \toprule
            \parbox{1.0cm}{\centering $P$\\{\footnotesize(\,pass./h)}} & \parbox{1.8cm}{\centering $U$\\{\footnotesize(bus-km/pass.)}} & \centering DBA & \parbox{1.0cm}{\centering $f_{\rm{R1}}$\\{\footnotesize (bus/h)}} & \parbox{1.0cm}{\centering $f_{\rm{R3}}$\\{\footnotesize (bus/h)}} & \parbox{1.0cm}{\centering $f_{\rm{R5}}$\\{\footnotesize (bus/h)}} & \parbox{1.0cm}{\centering $f_{\rm{R9}}$\\{\footnotesize (bus/h)}} & \parbox{1.5cm}{\centering $C_{\rm T}$\\{\footnotesize (bus-h)}} & \parbox{1cm}{\centering $C_{\rm U}/C_{\rm O}$} & \parbox{1cm}{\centering $v_{\rm p}$\\{\footnotesize (km/h)}} & \parbox{1cm}{\centering $\Delta C_{\rm{T}}$}\\
            \midrule
            \multirow{9}{*}{$4.0\times 10^4$} & \multirow{3}{*}{0.4} & \centering A & \textbf{32} & \textbf{47} & 47 & 32 & 3361(4) & 2.161(5) & 41.77(9) & --\\ \cline{3-11}
            & & \centering{B} & 32 & \textbf{47} & \textbf{47} & 32 & {3366(3)} & {2.161(5)} & {41.70(9)} & 0\% \\ \cline{3-11}
            & & \centering{C} & 31 & \textbf{46} & \textbf{46} & 31 & {3372(4)} & {2.242(4)} & {41.16(7)} & 0.3(2)\% \\ \cline{2-11}
           
            & \multirow{3}{*}{1.2} & \centering {A} & \textbf{53} & \textbf{53} & 53 & 35 & {7714(9)} & {4.767(8)} & {45.16(7)}  & -- \\ \cline{3-11}
           & & \centering{B} & 49 & \textbf{49} & \textbf{49} & 49 & {7770(10)} & {4.79(1)} & {44.78(9)}  & 2.2(3)\% \\ \cline{3-11}
           & & \centering{C} & 46 & \textbf{46} & \textbf{46} & 46 & {7850(10)} & {5.223(9)} & {43.72(7)} & 10.4(8)\% \\ \cline{2-11}

            & \multirow{3}{*}{2.0} & \centering {A} & \textbf{54} & \textbf{54} & 54 & 54 & {11940(30)} & {7.16(2)} & {45.8(1)} & -- \\ \cline{3-11}
           & & \centering{B} & 50 & \textbf{50} & \textbf{50} & 50 & {12050(20)} & {7.84(2)} & {44.92(8)} & 2.4(3)\%\\ \cline{3-11}
           & & \centering{C} & 67 & \textbf{44} & \textbf{44} & 44 & {12230(10)} & {7.732(9)} & {44.31(7)} & 14.4(7)\% \\\hline\hline

            \multirow{9}{*}{$5.0\times 10^4$} & \multirow{3}{*}{0.4} & \centering {A} & \textbf{39} & \textbf{58} & 58 & 39 & {4168(6)} & {2.171(5)} & {42.04(9)} & -- \\ \cline{3-11}
            & & \centering{B} & 39 & \textbf{39} & \textbf{58} & 39 & {4230(5)} & {2.553(5)} & {39.48(7)} & 14.7(4)\% \\ \cline{3-11}
            & & \centering{C} & 38 & \textbf{38} & \textbf{56} & 38 & {4251(7)} & {2.673(7)} & {38.8(1)} & 27.2(5)\% \\ \cline{2-11}
           
            & \multirow{3}{*}{1.2} & \centering {A} & \textbf{41} & \textbf{62} & 62 & 41 & {9800(20)} & {5.96(2)} & {42.9(1)} & -- \\ \cline{3-11}
           & & \centering{B} & 58 & \textbf{39} & \textbf{58} & 58 & {10080(10)} & {5.862(9)} & {41.78(6)} & 21.9(5)\% \\ \cline{3-11}  
           & & \centering{C} & 55 & \textbf{36} & \textbf{55} & 55 & {10200(10)} & {6.372(9)} & {40.83(6)} & 39.2(4)\% \\ \cline{2-11}

            & \multirow{3}{*}{2.0} & \centering {A} & \textbf{50} & \textbf{50} & 75 & 50 & {15320(20)} & {9.00(2)} & {43.51(7)} &  -- \\ \cline{3-11}
           & & \centering{B} & 97 & \textbf{49} & \textbf{49} & 49 & {15790(40)} & {7.97(3)} & {42.7(1)} & 22.4(4)\% \\ \cline{3-11}
           & & \centering{C} & 55 & \textbf{36} & \textbf{55} & 55 & {16080(20)} & {10.62(1)} & {40.83(6)} & 43.8(6)\% \\
            \bottomrule
        \end{tabular}\label{best_freq}
        \end{table*}
        \begin{figure}[t]
          \centering
          \includegraphics[width=0.45\textwidth]{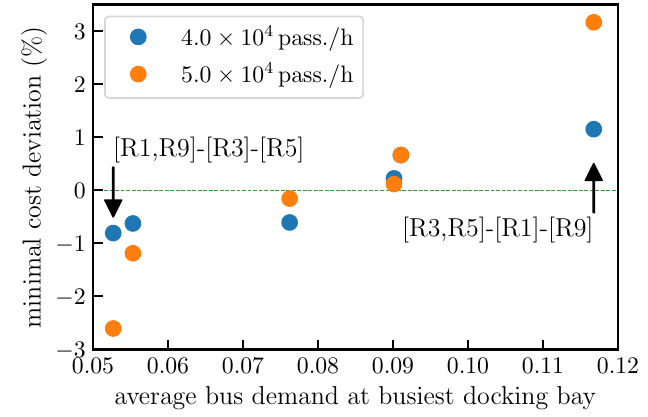}
          \caption{Correlation between the minimal total cost and the average bus demand at the busiest docking bay for all DBA with two services sharing the docking bay 1. The total cost is represented as a deviation from the average. For each DBA, the busiest docking bay is the one with the largest effective passenger traffic among the main hubs. The results are presented for passenger demands $4.0\times 10^4\,\mathrm{pass./h}$ and $5.0\times 10^4\,\mathrm{pass./h}$. The user cost factor has been set to $U=2.0\,\rm{bus\mbox{-}km/pass.}$. \label{cost_demand}}
          \end{figure}

        In the following, we focus our analysis on docking bay assignments (DBA) with $N_{\rm{max}}=2$. These DBA make use of all the available docking bays and would be preferred in practical operation. We have shown that when all bus services have the same frequency different optimal service frequencies and different optimal total costs are observed, even among the DBA with $N_{\rm{max}}=2$. If extrapolated to any service frequency configuration, this observation suggests that the solutions to the frequency optimization problem in a BRT system might depend on the DBA at the busiest stations.

        To further investigate on that possibility, we have approximated the frequency optimization problem by performing frequency scans for all possible relative frequencies \{$N_\text{1}$, $N_\text{3}$, $N_\text{5}$, $N_\text{9}$\}, with $N_i\in\text{\{1,2,3\}}$. For a given value of the user cost factor, $U$, the total cost has been determined for all the studied service frequency configurations. The best service frequency configuration has been identified as the one that yields the minimal total cost. 
        
        This analysis was initially performed for three different DBA: [R1,R3]-[R5]-[R9], [R3,R5]-[R1]-[R9] and [R1]-[R9]-[R3,R5]. While all these DBA make use of all the available docking bays, they are expected to behave differently because they combine either the R1 and R3 or the R3 and R5 services in the same docking bay. These two combinations exhibited the largest and smallest critical frequencies in our previous analysis, respectively. By comparing the results for configurations [R3,R5]-[R1]-[R9] and [R1]-[R9]-[R3,R5] we can, in addition, investigate the effects of the location of the shared docking bay on the best frequency configurations.

        The obtained best service frequency configurations for different operating conditions are summarized in Table~\ref{best_freq}. The Table also shows the total cost, $C_{\rm{T}}$, the user cost to operation cost ratio, $C_{\mathrm{U}}/C_{\mathrm{O}}$, and the average passenger speed, $v_{\mathrm{p}}$, in the best frequency configuration. As expected, as the user cost factor increases, i.e., as more importance is given to the passenger time, larger service frequencies and larger average passenger speeds are observed in the best frequency configurations. The ever-increasing passenger speeds suggest that in the best frequency configuration bus queues are always avoided.
        
        Notably, only in the case where both the user cost factor and the passenger demand are the smallest do the three studied DBA yield a similar set of the best service frequencies. As either the passenger demand or the user cost factor increase, sizable differences in the best service frequency configurations appear, even when only the location of the shared docking bay is changed. This is a remarkable result that suggests that changes in either the services sharing the docking bay, or the physical location of the shared docking bay can significantly affect the optimal service frequencies in a BRT system.
        
        This finding is further confirmed by the large sensitivity of the system to changes in the DBA at the main hubs, as evidenced in the last column of Table~\ref{best_freq}, $\Delta C_{\rm T}$. This column exhibits the expected change in the total cost of the system if, while using the best frequencies obtained for DBA [R1,R3]-[R5]-[R9] (noted as A), the DBA at the main hubs were changed to [R3,R5]-[R1]-[R9] (noted as B) or [R1]-[R9]-[R3,R5] (noted as C). Once again, only in the case with the smallest passenger demand and user cost factor the performance of the system would not be affected. As these parameters increase, increments up to 43.8\% in the total cost are expected upon changes in the DBA at the main hubs. Although this is an extreme case, it becomes evident that the frequency optimization problem is not independent of the docking bay assignment at the busiest stations in the system and that significant performance losses could be observed if this correlation is ignored, especially in BRT systems with a large passenger demand.

        We remark that the values of $C_{\mathrm{U}}/C_{\mathrm{O}}$ in the best service frequency configurations shown in Table~\ref{best_freq} are in all cases smaller than 11, which confirms that the observed differences among the studied DBA are not the consequence of unrealistically high user cost factors.
        
        To further learn about the influence of the DBA on the performance of a BRT system, let us now focus on the minimal values of the total cost. By comparing the results for DBA [R3,R5]-[R1]-[R9] and [R1]-[R9]-[R3,R5] in Table~\ref{best_freq}, it becomes evident that changing the shared docking bay from 1 to 3 (3 to 1 for west-bound services) leads to an increment in the total cost of the system and a reduction in the average passenger speed in the best configuration. This observation is an extension of our previous finding, according to which larger critical frequencies are obtained when the shared docking bay is set to be the number 1 (number 3 for west-bound services). Based on this analysis we can conclude that the busiest docking bay in a BRT should be the first one buses encounter on their arrival in the station. A result that could be readily applied on operating BRT systems.

        Nevertheless, even between DBA [R1,R3]-[R5]-[R9] and [R3,R5]-[R1]-[R9], respectively noted as A and B in Table~\ref{best_freq}, there are differences in the total cost and the average passenger speed in the best frequency configurations. To further understand the origin of these differences we have extended our approximated frequency optimization problem to all DBA where two services share the docking bay number 1 (number 3 for west-bound services). The results obtained using a user cost factor $U=2.0\,\rm{bus\mbox{-}km/pass.}$ are presented in Figure~\ref{cost_demand}, where the variation of the minimal total cost for each DBA with respect to the average is presented as a function of the average bus demand at the busiest docking bay. The busiest docking bay is the one with the largest passenger traffic among the main hubs. We notice that significant variations up to 2.0\% and 5.0\% are observed in the minimal total cost among the studied DBA when the passenger demand is set to $P=4.0\times 10^5\,\textrm{pass./h}$ and $P=5.0\times 10^5\,\textrm{pass./h}$, respectively. As evident from the Figure, this result can be explained in terms of the average bus demand at the busiest docking bay. 
        
        Notably, the DBA yielding the smallest minimal cost is [R1,R9]-[R3]-[R5], which according to Panel (c) of Figure~\ref{ODmatrix} yields the largest effective passenger traffic. Therefore, to determine the best possible DBA in a BRT system, the expected bus occupation at the busiest docking bays is more important than the expected passenger accumulation in the station. If the bus demand is small, even if the passenger traffic is large, buses are more capable to collect passengers and smaller frequencies are needed to satisfy the demand, queues are thus less likely to form.
        
        In summary, we have shown that the set of optimal service frequencies in a BRT system strongly depends on the docking bay assignment at the busiest stations. In addition, we have shown that the performance of the BRT system can be improved if the shared docking bay is the first one buses encounter on arrival at the station, and if the average bus demand in the busiest docking bay is minimized.
       
    \section{Conclusions}
        We have implemented a cellular automaton (CA) microsimulation scheme for the entire operation of a simplified BRT system, including both bus movement and passenger transportation. This approach offers the advantage of properly modeling bus queue formation and passenger accumulation at the stations. 

        The evolution of the BRT system's performance parameters with the service frequency shows that changes in the docking bay assignment (DBA) at the main hubs introduce significant changes in the performance of the system. In particular, the critical frequency, above which bus queues start to form, directly depends on the maximal number of services sharing a docking bay, the actual services sharing the docking bay and the physical location at the station of the docking bay being shared.

        Because of the differences in the critical frequency among different DBA, the service frequencies minimizing the total cost are also dependent on how the bus services are set to stop at the busiest stations. This correlation is a consequence of the limitations imposed to the optimal frequencies by the critical frequency. We showed that, even when all the available docking bays are used, the mere change of the DBA at the main hubs can lead to significant increments in the total cost of the system, in the worst identified scenario the increment reached 43.8\%. Even though in the optimal configuration of a BRT system queues should not be present, considering the bus interactions is crucial to determine the upper bounds for the service frequencies for each DBA. 

        In general, we observed that the performance of the system is enhanced if the busiest docking bay is the first one that buses encounter when reaching a station. Further performance improvements can be achieved if the average bus demand for the services stopping at the busiest docking bay is minimized. 

        From our results it can be concluded that the frequency optimization problem should consider bus interactions, the bus demand, and the passenger traffic. It should also include the docking bay assignment, at least at the busiest stations, as an additional operational variable. This is particularly important for crowded BRT systems where bus and passenger queues appear regularly.

        By properly tuning the parameters that influence bus and passenger dynamics, our framework has the potential to accurately reproduce realistic BRT system behavior. It is therefore a useful option for system operators to evaluate their performance and evaluate the impact of operation changes.

\section{Author Contributions}
The authors confirm contribution to the paper as follows: study conception and design and data collection: MAUL; analysis, interpretation of results and draft manuscript preparation: MAUL and WFOP. All authors reviewed the results and approved the final version of the manuscript.

\section{Declaration of Conflicting Interests}
The authors declared no potential conflicts of interest with respect to the research, authorship, and/or publication of this article.

\begin{acks}
  This work has been partially carried out with resources provided by the CYTED cofunded Thematic Network RICAP (517RT0529), and has been funded by Universidad de La Sabana under grant numbers ING-196-2017 and ING-273-2021.
\end{acks}

\end{document}